\documentclass{aastex}
\usepackage{spr-astr-addons}
\usepackage{url}
\usepackage{graphicx}
\usepackage{epsfig}
\usepackage{bm}
\usepackage{amsfonts}
\usepackage{subfigure}
\usepackage{subfigure}

\RequirePackage{color}

\begin{document}

\title{Effects of anisotropy on   interacting ghost dark energy in Brans-Dicke theories }
\slugcomment{Not to appear in Nonlearned J., 45.}
\shorttitle{Short article title}
\shortauthors{Autors et al.}

\author{H. Hossienkhani\altaffilmark{1}}
\altaffiltext{1}{Department of Physics, Hamedan Branch, Islamic Azad University, Hamedan, Iran.}
\author{V. Fayaz\altaffilmark{2}}
\altaffiltext{1}{Department of Physics, Hamedan Branch, Islamic Azad University, Hamedan, Iran.}
\author{N. Azimi \altaffilmark{3}}
\altaffiltext{3}{Department of Mathematics, Hamedan Branch, Islamic Azad University, Hamedan, Iran.}
\affil{hossienhossienkhani@yahoo.com}
\affil{$fayaz_{-}vahid@yahoo.com$}
\affil{Azimi1379@yahoo.com}
\begin{abstract}
By  interacting   ghost dark energy (ghost DE)   in the framework of  Brans-Dicke theory, a spatially homogeneous and anisotropic Bianchi type I Universe has been studied.
 For this purpose, we use the squared sound speed $c_s^2$ whose sign determines the stability of the model. As well as  we
probe observational constraints on the ghost dark energy models as the unification of dark matter and dark energy by using the latest observational data.  In
order to do so,    we focus on observational determinations of the  Hubble expansion rate (namely, the expansion history) $H(z)$. After that we evaluate
the evolution of  the growth of perturbations in the linear regime for both ghost DE and   Brans-Dicke theory
  and compare the results with standard FRW and $\Lambda$CDM models.  We  display the effects of the   anisotropy on the evolutionary behaviour the ghost DE models
 where  the growth rate is higher in this models.  Eventually  the growth factor    for the  $\Lambda$CDM Universe   will always fall behind the  ghost DE models in an anisotropic Universe.

\end{abstract}

\keywords{ Anisotropic Universe, Holographic dark energy, Brans-Dicke theories, Stability of theory}


\section{Introductions}
Accelerating expansion of the Universe \citep{1,2} can be demonstrated either by a missing energy element which can be usually called ``dark energy" (DE) with an exotic equation of state (EoS), or by modifying
the underlying theory of gravity on large scales.  The other models of DE  have been discussed widely in  literature  considering a cosmological constant \citep{4}, a canonical scalar field (quintessence) \citep{5}, a phantom field, which is a scalar field with a negative sign of the kinetic term \citep{6,7}, or the combination of quintessence and phantom in a unified model named quintom \citep{8}.
Among various models of DE, a new model of DE called Veneziano ghost dark energy (ghost DE) of our interest has been  suggested recently  \citep{9,10}.
It is supposed to exist to solve the $U(1)$ problem in low-energy effective theory of QCD, has attracted a lot of interests in recent years   \citep{11,12,14,15}, though it is totaly decoupled from the physical sector \citep{15}. There are some DE models where the ghost plays the role of DE (see, e.g., \citep{16}) and becomes a real propagating physical degree of freedom subjected to some severe constraints. They have  explored a DE model with a ghost scalar field in the context of the runaway dilaton scenario in low-energy   effective string theory and addressed the problem of vacuum stability by implementing higher-order derivative terms  and shown that a cosmologically viable model of "phantomized" DE can be constructed without violating the  stability of quantum fluctuations.  Nevertheless,  the Veneziano ghost is not a physical propagating degree of freedom and the corresponding GDE model does not violate unitarity causality or gauge invariance and other crucial features of renormalizable quantum field theory, as advocated in \citep{17,18,19}.\\
  Scalar-tensor    theory provide the most natural generalizations of General Relativity (GR) by presenting additional fields. In this theory, the field
equations are even more complex then in GR. One   the simplest of    the scalar tensor is the Brans-Dicke (BD) theory of gravity which was proposed by  \citep{20}. BD theory involves a scalar field and is perhaps the most viable alternative theory to Einstein's general theory. It also passed
the observational tests in the solar system domain \citep{21}. Since the  ghost DE  model have a dynamic behavior it is more reasonable to consider this model in a dynamical framework such as BD theory. It was shown that some features of original ghost DE in BD cosmology differ from Einstein's gravity \citep{22}. For example while the original ghost DE is instable in all range of the parameter spaces in standard cosmology \citep{23}, it leads to a stable phase in BD theory \citep{24}.\\
 Recent experimental data such as theoretical arguments support the existence of anisotropic expansion phase, which evolves into an isotropic one. It forces one to study evolution of the Universe with the anisotropic background.   The possible effects of anisotropy in the early
Universe have been studied with Bianchi  type I (BI) models from different points of view  \citep{25,26,27,27a,27b,27c}.  Recently, \citep{28} importance of BI model  have shown to discuss the effects of anisotropy on the basis of recent evidences. Some exact anisotropy  solutions have  been also investigated in this BD theory \citep{28a,28b,28c}.  Lately, Hossienkhani \citep{29} investigated the interacting DE scalar fields models in an anisotropic Universe. Consequently, it would be worthwhile to explore anisotropic DE models in the context of BD theory.  In this work we study the  evolution of  the Hubble parameter,    squared sound speed and    growth of perturbations in ghost DE of BD theory. The  ghost DE  model is considered as a dynamical DE model with  varying EoS parameter which can dominate the Hubble flow and influence the growth of structure in the Universe. Here we consider the interacting case of  ghost DE model in BI model.\\
This paper is organized as follows. In the next section,  we review the interacting  ghost DE model in an anisotropic Universe and describe the evolution of background
cosmology in this model.  In section 3 we discuss the linear evolution of perturbations in ghost DE cosmology of BI model.
Sect. 4, we formulate the field equations of BD theory for BI Universe and provide the solution to the
field equations with interaction  between DM and DE.   Finally we conclude in  Sect. 5.
\section{Metric and ghost dark energy model}
We consider a class of homogeneous and anisotropic models where the line component is of the Bianchi type I,
\begin{equation}\label{1}
ds^2=dt^{2}-A^{2}(t)dx^{2}-B^{2}(t)dy^{2}-C^{2}(t)dz^{2},
\end{equation}
with  $A, B, C$ being the functions of time only.  This model is an anisotropic generalization of the Friedmann
model  with Euclidean spatial geometry.  Note that the Kantowski-Sachs (KS)   is recovered when one takes  $B=C$.  The contribution of the interaction with the matter fields is given by the energy momentum tensor which, in this case, is defined as
\begin{eqnarray}\label{2}
T^{\mu}_{\nu}=diag[\rho,-\omega\rho,-\omega\rho,-\omega\rho],
\end{eqnarray}
where $\rho$ and $\omega$ describe the energy density and EoS parameter respectively. By taking a preferred timelike vector field (four velocity) $u^i$, which
satisfies $u^iu_i=1$,   we can write the following   Einstein's field equations  for BI  model  \citep{30}:
\begin{eqnarray}
&&3H^{2}-\sigma^{2}=\frac{1}{M_p^2}(\rho_{m}+\rho_{\Lambda}),  \label{3} \\
&&3H^2+2\dot{H}+\sigma^{2}=-\frac{1}{M_p^2}\left(p_{m}+p_{\Lambda}\right), \label{4}  \\
&&R=-6(\dot{H}+2H^2)-2\sigma^2,\label{5}
\end{eqnarray}
where  $M_p^2=1/(8\pi G)$, $\rho_{\Lambda}$ and $p_{\Lambda}$ are the Planck mass, the energy density and pressure of dark energy, respectively, $a=(ABC)^{\frac{1}{3}}$ is the average scale factor, and
$\sigma^2=1/2\sigma_{ij}\sigma^{ij} $  in which
$\sigma_{ij}=h_i^\gamma u_{(\gamma;\delta)}h_j^\delta+\frac{1}{3}\theta h_{ij}$
is the shear tensor, which describes the rate of distortion of the
matter flow,  $\theta=3H=u^{j}_{;j}$ is the  scalar expansion and $h_{ij}$ is the
projection tensor defined from the expression $h_{ij}=g_{ij}+u_{i}u_{j}$.  It may be pointed out here  if one sets $\sigma=0$, the equations reduce to that
obtained for a flat FRW Universe. Therefore when the Universe is sufficiently large it
behaves like a flat Universe.  Let us take  the average Hubble parameter  and the shear scalar $\sigma^2$ as
\begin{eqnarray}
H&=&\frac{\dot{a}}{a}=\frac{1}{3}\bigg(\frac{\dot{A}}{A}+\frac{\dot{B}}{B}+\frac{\dot{C}}{C}\bigg),   \label{6} \\
2\sigma^{2}&=&\sigma_{\mu\nu}\sigma^{\mu\nu}=\left(\frac{\dot{A}}{A}\right)^{2}+
\left(\frac{\dot{B}}{B}\right)^{2}+\left(\frac{\dot{C}}{C}\right)^{2} -\frac{\theta^{2}}{3}. \label{7}
\end{eqnarray}
We investigate the ghost DE model in the framework of Einstein gravity. The ghost DE density is given by \citep{10,31}
\begin {equation}\label{8}
\rho_{\Lambda}=\alpha H,
\end{equation}
where $\alpha$ is a constant with dimension $[energy]^3$, and roughly of order of $\Lambda_{QCD}^3$ where $\Lambda_{QCD} \sim100MeV$.
Using  (\ref{3}),   the dimensionless density parameter can also be defined as usual
\begin{eqnarray} \label{9}
\Omega_m =\frac{\rho_m}{\rho_{cr}}, \quad   \Omega_{\Lambda} =\frac{\rho_{\Lambda}}{\rho_{cr}}=\frac{ \alpha}{3M_p^2H},
\end{eqnarray}
where the critical energy density is $\rho_{cr}=3M_p^2H^2$.  By using Eq. (\ref{9}), the first BI   (\ref{3}), can be written as
\begin{equation}\label{10}
\Omega_m+\Omega_{\Lambda}=1-\frac{\sigma^2}{3H^2} .
\end{equation}
We shall take that the shear scalar can be described based on the average Hubbel parameter, $\sigma^2=\sigma_0^2H^2$, where $\sigma_0^2$ is a constant. So, Eq. (\ref{10}) lead to
\begin{eqnarray}\label{11}
\Omega_m+\Omega_{\Lambda}=1-\Omega_{\sigma0},\quad ~with ~~\Omega_{\sigma0}=\frac{\sigma_0^2}{3},
\end{eqnarray}
where $\Omega_{\sigma0}$ is the anisotropy parameter. For inserting   the energy density of the DE component, we use Eq. (\ref{8})  into  (\ref{3}) in order to obtain the Hubble parameter in ghost DE cosmologies
\begin{eqnarray}\label{12}
&&H=\sqrt{(\frac{\alpha}{6M_p^2(1-\Omega_{\sigma 0})})^2+\frac{\rho_{m0}a^{-3}}{3M_p^2(1-\Omega_{\sigma 0})}}\cr
&&~~~~~~~~~~~~~~~~~~~~~~~~~~~~~~~~+\frac{\alpha}{6M_p^2(1-\Omega_{\sigma 0})}.
\end{eqnarray}
In terms of the dimensionless energy density $\Omega_{mo}=\rho_{m0}/(3 H_0^2M_p^2)$   and redshift parameter $z=1/a-1$, the above Hubble equation becomes
\begin{eqnarray}\label{13}
&&H=\frac{H_0}{2(1-\Omega_{\sigma 0})}\bigg[1-\Omega_{m0}-\Omega_{\sigma 0}\cr
&&+\sqrt{(1-\Omega_{m0}-\Omega_{\sigma 0})^2+4 \Omega_{m0}(1-\Omega_{\sigma 0})(1+z)^{3}}\bigg].
\end{eqnarray}
In the $\Lambda$CDM model Hubble's parameter is $H=H_0( \frac{\Omega_{m0}(1+z)^{3} + \Omega_\Lambda}{1-\Omega_{\sigma 0}})^{\frac{1}{2}} $ and the EoS of  DE is fixed to be $\omega_\Lambda=-1$. For model such as $w$CDM (with the constant EoS $w$), it is \\$H=H_0(  \frac{\Omega_{m0}(1+z)^{3}+ (1- \Omega_{m0}-\Omega_{\sigma 0})(1+z)^{3(1+w)} }{1-\Omega_{\sigma 0}})^{\frac{1}{2}}$. The Hubble constant $H_0$ in    (\ref{13})   is taken as $72~km/s~Mp/c$, in the whole discussion.  Another   the
Hubble constant measurements, $H_0= 73.8 \pm 2.4~  kms^{-1}Mpc^{-1}$  from \citep{32},  $H_0=73  \pm 3 ~kms^{-1}Mpc^{-1}$ from the combination WMAP \citep{33}, and the other with $H_0= 68 \pm   4 ~ kms^{-1}Mpc^{-1}$ from a median statistics analysis of 461 measurements of $H_0$ \citep{34,35}. The conservation equations for pressureless dust matter and DE in the presence of interaction are
\begin{eqnarray}\label{14}
\dot{\rho}_\Lambda+3H\rho_\Lambda(1+\omega_\Lambda)&=&-Q,\\
\dot{\rho}_m+3H\rho_m&=&Q,\label{15}
\end{eqnarray}
where the dot is the derivative with respect to cosmic time,  $\omega_\Lambda$ is the DE EoS parameter and $Q$ stands for the interaction term. Following  \citep{36,37},   we shall assume $Q=3b^2H(\rho_m+\rho_\Lambda)$ with the coupling constant  $b^2$.  Differentiating Eq. (\ref{3})  with respect to time, we
obtain
\begin{eqnarray}\label{16}
\frac{d{H}}{dz}&=&\frac{3H}{2(1+z)}\frac{\Omega_{\Lambda}(z)}{1-\Omega_{\sigma 0}}\bigg(1+r+\omega_{\Lambda}(z)\bigg),\cr
  r&=&\frac{1-\Omega_{\Lambda}(z)-\Omega_{\sigma 0}}{\Omega_{\Lambda}(z)}.
\end{eqnarray}
Combining Eqs. (\ref{8}) and (\ref{16}) with the continuity equation given in Eq. (\ref{14}),  the EoS parameter for ghost DE   model is
\begin{equation}\label{17}
\omega_{\Lambda}(z)=\frac{1-\Omega_{\sigma0}}{-2+\Omega_{\Lambda}(z)+2\Omega_{\sigma0}}\left(1+\frac{2b^2}{\Omega_{\Lambda}(z)}(1-\Omega_{\sigma0})\right).
\end{equation}
One can easily check that in the late time where $\Omega_{\Lambda}\rightarrow1$ and $\Omega_{\sigma0}\rightarrow0$, the EoS parameter of interacting ghost DE necessary crosses the phantom line, namely, $\omega_{\Lambda}=-(1+2b^2)<-1$ independent of the value of coupling constant $b^2$.  For present time with taking  $\Omega_{\sigma0}=0.001$ and $\Omega_{\Lambda}^0=0.69$, the phantom crossing can be achieved provided $b^2>0.1$. We now
calculate the equation of motion for the energy density of DE in ghost DE model. Taking the time derivative of $\Omega_{\Lambda}$  in Eq. (\ref{9}) and  using relation $\dot{\Omega}_{\Lambda}=-H(z)(1+z)\Omega_{\Lambda}'(z)$,  we obtain
\begin{equation}\label{18}
\Omega_\Lambda'(z)=-\frac{3\Omega_{\Lambda}(z)}{1+z}\bigg(\frac{-1+\Omega_{\sigma0}+\Omega_{\Lambda}(z)+b^2(1-\Omega_{\sigma0})}{-2+\Omega_{\Lambda}(z)+2\Omega_{\sigma0}}\bigg),
\end{equation}
where prime means differentiation with respect to the redshift $z$. We can determine the deceleration parameter $(q)$ as  $q=-1+\frac{1+z}{H}\frac{dH}{dz}$ as follow.  Using Eqs. (\ref{16}),   (\ref{17}) and in the presence of interaction the deceleration parameter is obtained by
\begin{equation}\label{19}
q(z)=\frac{1}{2}+\frac{3}{2}\left(\frac{\Omega_{\Lambda}(z)+2b^2(1-\Omega_{\sigma0})}{-2+\Omega_{\Lambda}(z)+2\Omega_{\sigma0}}\right),
\end{equation}
where $\Omega_{\Lambda}(z)$ is given by Eq. (\ref{18}). The speed of sound $c^2_s$ is defined as \footnote{ In classical perturbation
theory we assume a small fluctuation in the background of the
energy density and we want to observe whether the perturbation
will grow or fall. In the linear perturbation factor, the perturbed energy density   is  $\rho(t,x)=\rho(t)+\delta\rho(t,x)$, where $\rho(t)$ is the unperturbed   energy density. The energy conservation equation $ T^{\mu\nu}_{;\mu}$ yields  $\delta\ddot{\rho}=c_s^2 \nabla^2 \delta \rho(t,x)$ \citep{4}, where $c^2_s=dp/d\rho$ is the square of the sound speed.}
\begin{equation}\label{20}
c_s^2 =\frac{\dot{p}_\Lambda}{\dot{\rho}_\Lambda}=\frac{\rho_\Lambda}{\dot{\rho}_\Lambda}\dot{\omega}_\Lambda+\omega_\Lambda.
 \end{equation}
Now by computing $\dot{\omega}_\Lambda$ and using  Eqs.   (\ref{8}),  (\ref{14}) and  (\ref{17}) which reduces to
\begin{eqnarray}\label{21}
&&c_s^2 =\frac{2(-1+\Omega_{\sigma0})}{\Omega_\Lambda(-2+2\Omega_{\sigma0}+\Omega_\Lambda)^2}\bigg(\Omega_\Lambda(1-\Omega_{\sigma0}-\Omega_\Lambda) \cr
&&~~~~~~~~~~~+b^2(\Omega_{\sigma0}-1)(-4+4\Omega_{\sigma0}+3\Omega_\Lambda)\bigg),
 \end{eqnarray}
  \begin{figure}[tb]
\includegraphics[width=.35\textwidth]{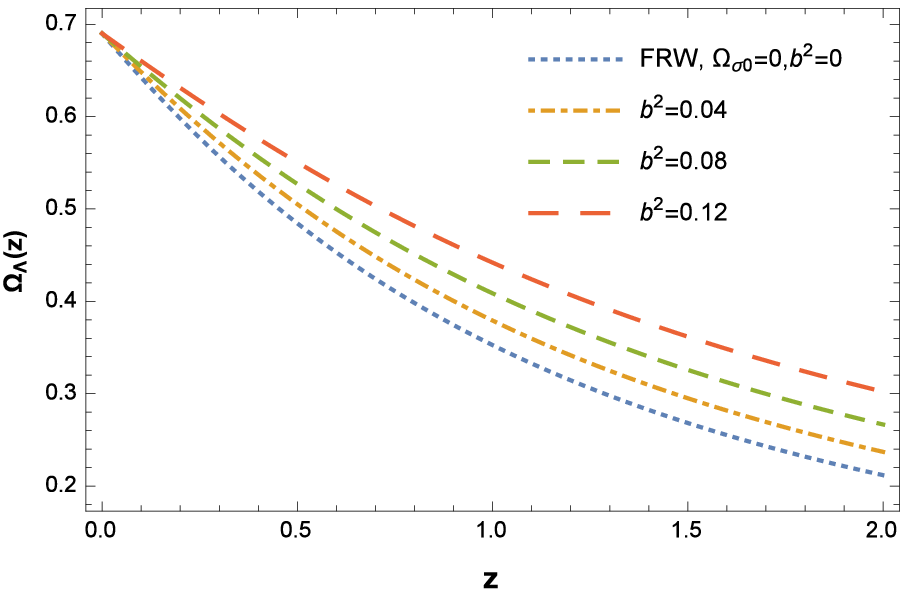}\hspace{.1cm}
\includegraphics[width=.35\textwidth]{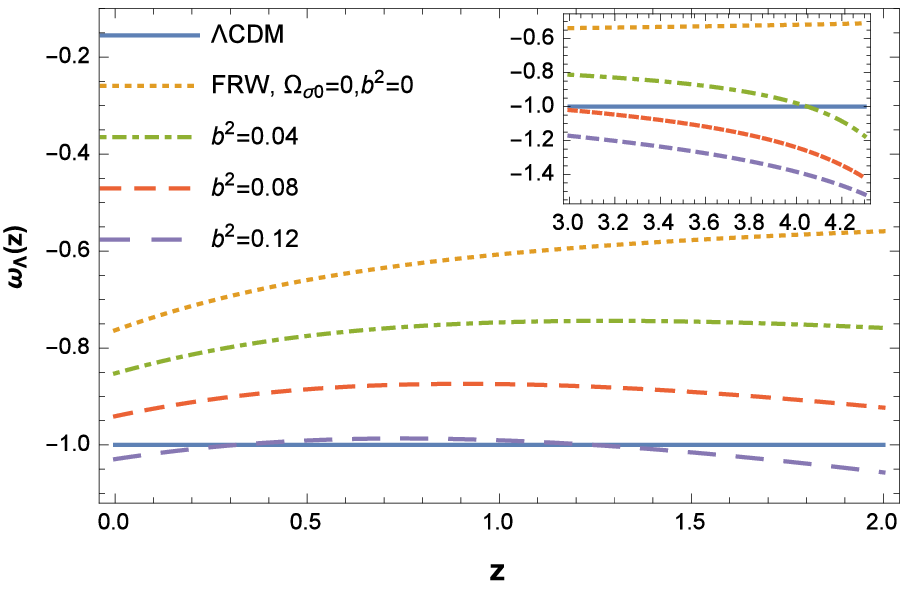}\hspace{.1cm}
\includegraphics[width=.35\textwidth]{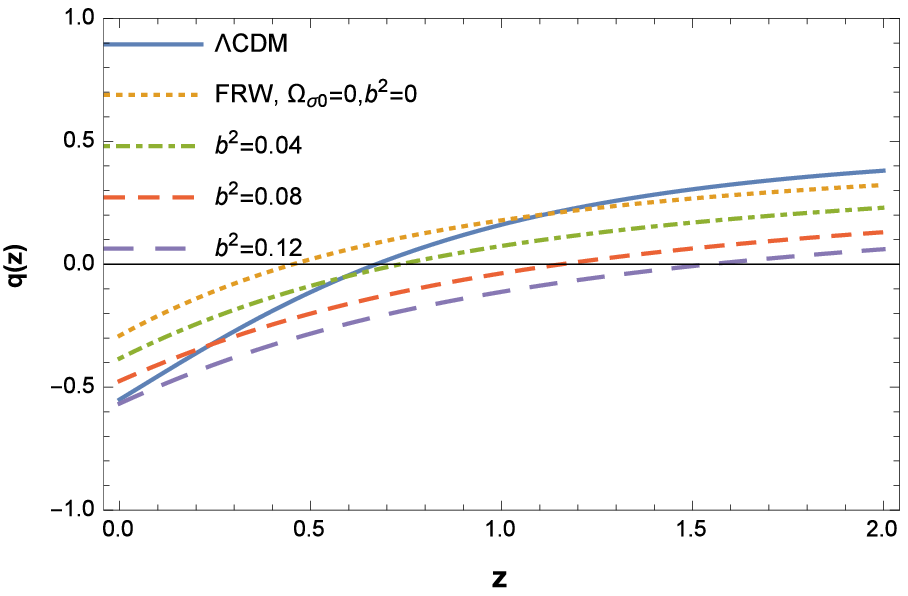}
\caption{Upper panel: The redshift evolution of the density parameters $\Omega_\Lambda(z)$. Middle panel: The evolution of EoS parameter $\omega_\Lambda(z)$.  Lower panel:  The  deceleration parameter $q(z)$  as a function of cosmic redshift $z$ for the different parameter $b^2$. Here we take  $\Omega_{\Lambda}^0=0.69$ and $\Omega_{\sigma0}=0.001$.}
 \label{fig:1}
 \end{figure}
 \begin{figure}[tb]
 \includegraphics[width=.35\textwidth]{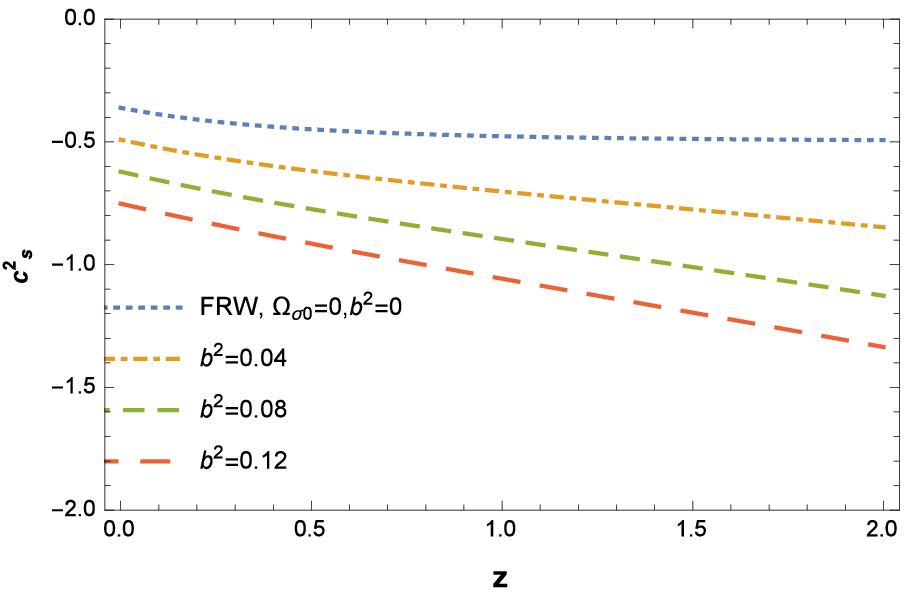}\hspace{.5cm}
 \includegraphics[width=.35\textwidth]{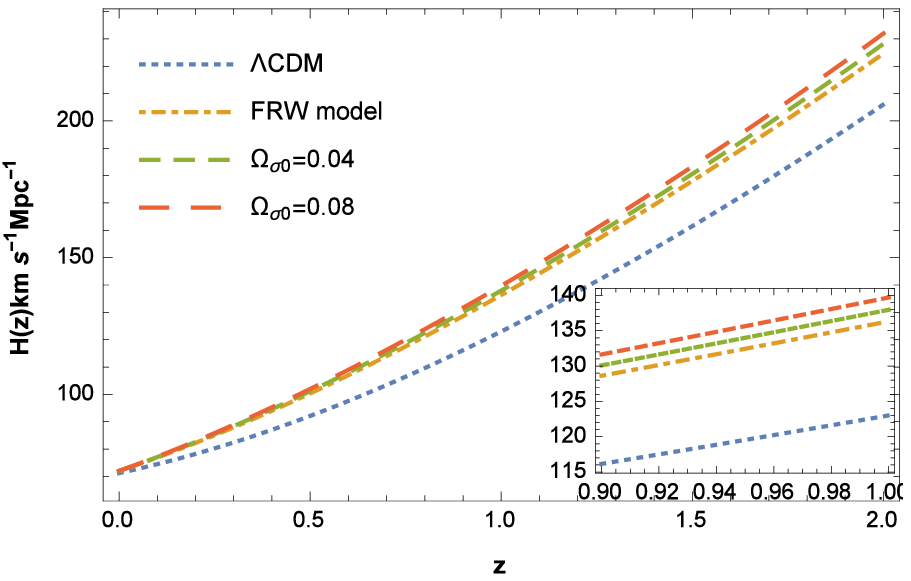}\hspace{.5cm}
 \includegraphics[width=.35\textwidth]{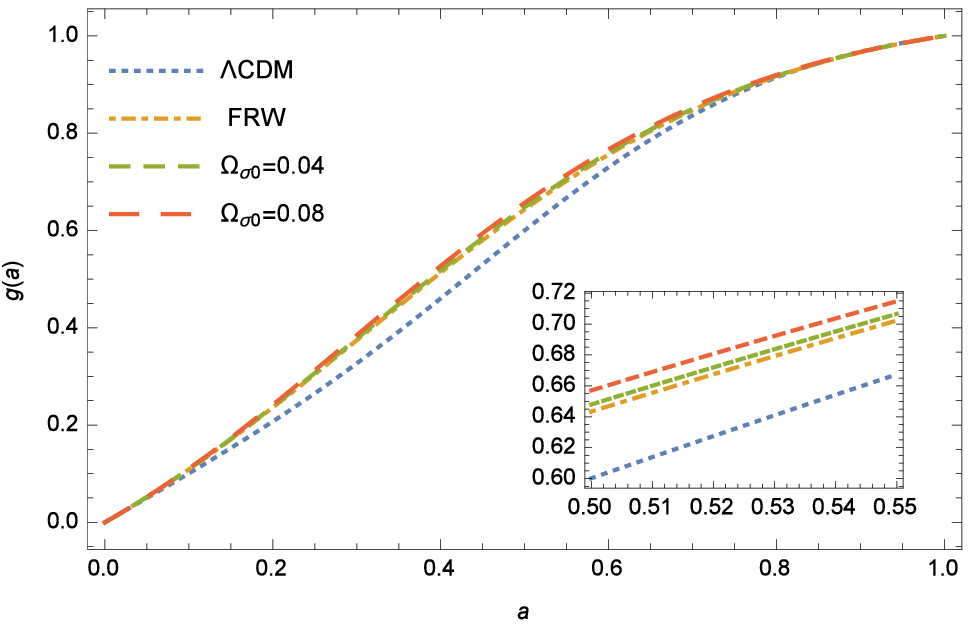}
\caption{Upper panel: The evolution of squared sound speed $c_s^2$ as a function of cosmic redshift $z$ for the different parameter $b^2$ with  $\Omega_{\Lambda}^0=0.69$ and $\Omega_{\sigma0}=0.001$.    Middle panel: Evolution of the  Hubble parameter $H(z)$  as a function of cosmic redshift $z$ for the different parameter $\Omega_{\sigma0}$ with $b^2=0.1$, $\Omega_{m0}=0.277$ and  $H_0=72kms^{-1}Mpc^{-1}$. Lower panel: Time evolution of the growth factor as a function of the scale factor for the different cosmological models and comparing to
the $\Lambda$CDM and FRW models.   Auxiliary parameters are the same as shown in the  middle panel of Fig. (2). }
 \label{fig:2}
 \end{figure}
which is the squared sound speed for interacting ghost DE fluid. It may be mentioned  that for causality and stability under perturbation it is required to satisfy the inequality condition $c_s^2\leq 1$ \citep{38}.\\
In Fig. (1)  we show the  energy density of DE component $\Omega_\Lambda$ (upper panel), the evolution of the EoS parameter $\omega_\Lambda$ (middle
panel), the  deceleration parameter $q(z)$ (lower panel) while in Fig. (2)  we show  the   squared sound speed $c_s^2$   (upper panel) and  Hubble parameter $H$
(middle panel) as a function of the cosmic redshift $z$ for   different values of the
model parameters $b^2$ and   $\Omega_{\sigma0}$,  and comparing to  FRW ghost DE and $\Lambda$CDM cosmological models.  In the case of
the   ghost DE model we have assumed the present values:  $\Omega_{\sigma0}=0.001$, $\Omega_{\Lambda}^0=0.69$ and  $H_0=72~km/s~Mp/c$. Also for the case  of
$\Lambda$CDM  model it  is    $\Omega_{\Lambda}^0=0.7$ and  $\Omega_{m0}=0.3$. From figure (1) we see that for the case of  $b^2\neq 0.12$,  the EoS parameter for ghost  DE model is always bigger than $\omega_\Lambda=-1$ and remains in the quintessence regime, i.e., $\omega_\Lambda>-1$  while for $b^2= 0.12$, we see that $\omega_\Lambda$ of the ghost DE can cross the phantom divide.  In the limiting case of the FRW Universe it was argued  \citep{42} that without interaction ($b^2=0$) $\omega_{\Lambda}$ is always larger than $-1$ and cannot cross the phantom divide while in the presence of the interaction the situation is changed.
Recent studies have constructed  $q(z)$ takeing into account that the strongest evidence of accelerations happens at redshift of  $z\sim0.2$. In order to  do this, they  have  set $q(z)=1/2 (q_1 z+q_2)/(1+z)^2$ to reconstruct it and after that they have obtained  $q(z)\sim-0.31$ by fitting this model  to the  observational data \citep{43}.
 Under these conditions and considering bottom panel of figure (1), we give the present value of the deceleration parameter
for the interacting  ghost DE with $b^2= 0.12$ is $q_0 \simeq -0.56$, is significantly smaller than  $q_0 \sim-0.54$ for the  $\Lambda$CDM cosmological model  \citep{44}, as expected (see also  figure (1)). Graphical analysis of $c_s^2$   shows that our theory could be unstable  in FRW and BI models  as shown in upper panel of Fig. (2). Furthermore, we would see that the non interacting ghost DE in FRW is more stable than the interacting ghost DE in an anisotropic Universe. It is also interesting to see how our models  when compared with recent measurements of the Hubble parameter performed with the $\Lambda$CDM model. This comparison is done in figure 2 (middle panel), where we plot the   evolution of $H(z)$ depends on the value of the  $\Omega_{\sigma0}$ parameter for the ghost DE and   $\Lambda$CDM model considered in this work.  It was observed that the Hubble parameter are bigger in these models compared to the   $\Lambda$CDM model.
Also,  we can see that for the biggest value, the $\Omega_{\sigma0}$ parameter is taken, and the biggest value of the Hubble expansion rate $H(z)$ is gotten.
\section{Linear perturbation theory in ghost DE }
The coupling between the dark components could significantly affect not only the expansion history of the Universe,
but also the evolutions of the density perturbations, which would change the growth history of cosmic structure.
The  linear growth of perturbations  for the large scale structures is derived from matter era,  by calculating the evolution of the growth factor $g(a)$ in ghost DE models and compare it with the solution found for the $\Lambda$CDM model. The differential equation for $g(a)$ is given by \citep{39,40,41}
\begin{equation}\label{22}
g''(a)+(\frac{3}{a}+\frac{E'(a)}{E(a)})g'(a)-\frac{3}{2}\frac{\Omega_{m0}}{a^5E^2(a)}g(a)=0,
\end{equation}
for the prime denoting the derivative with respect to $\ln a$ and $E(z)=H/H_0$ is  the evolution of dimensionless Hubble parameter.  For a non interacting  DE model, by using Eqs. (\ref{13}), (\ref{16}) and (\ref{18}), we solve numerically Eq. (\ref{22}) for studying the linear growth with  ghost DE   in an anisotropic Universe. After that we compare the linear growth in  the ghost DE   model  with the linear growths in
 the $\Lambda$CDM  and FRW models. To evaluate the initial conditions, since we are in the linear regime, we take that the linear growth factor has a power law solution,
$g(a)\propto  a^n$, with $n>1$, then the  linear growth should grow in time.  In  bottom panel of Fig. (2)  we show the evolution of the linear growth factor
$g(a)$ as a function of the scale factor.  In the ghost DE model with $\Omega_{\sigma0}\neq0$, the growth factor evolves
proportionally to the scale factor, as expected.  In the FRW model ($\Omega_{\sigma0}=0$), the growth factor evolves more slowly compared to the BI model because   the     FRW model dominates  in the late time Universe.  In the case of $\Lambda$CDM,  $g(a)$ evolves more slowly than in the  ghost DE of FRW model since the expansion of the Universe slows down the structure formation.
\section{Bianchi type I field equations and ghost dark energy in   Brans-Dicke theory   }
The BD theory with self-interacting potential is described by the action  \citep{45,46,47}
\begin{equation}\label{23}
S=\int d^4x\sqrt{g}\left(-\frac{1}{8\omega_0}\phi^2R+\frac{1}{2}g^{\mu\nu}\partial_{\mu}\phi\partial_{\nu}\phi+L_m\right),
\end{equation}
where $\omega_0$ represent the constant BD parameter   and $L_m$  the matter part of the Lagrangian. We have taken $8\pi G_0=c=1$. In particular we may expect that $\phi$ is spatially uniform, but varies slowly with time. The nonminimal coupling term $\phi^2R$ where $R$ is the Ricci scalar, replaces with the Einstein-Hilbert term $\frac{1}{G_N}R$  in such a way that $G_{eff}^{-1}=\frac{2\pi}{\omega_0}\phi^2$ where $G_{eff}^{-1}$ is the effective gravitational constant as long as the dynamical scalar field $\phi$ varies slowly. Using the principle of least action, we obtain the field equations
\begin{eqnarray}\label{24}
&&\phi G_{\mu\nu}=-8\pi T_{\mu\nu}^{m}-\frac{\omega_0}{
\phi}\left(\phi_{,\mu}\phi_{,\nu}-\frac{1}{2}g^{\mu\nu}\phi_{,\gamma}\phi^{,\gamma}\right)\cr
&&~~~~~~~~~~~~~~~~~~~~~~-\phi_{;\mu;\nu}+g_{\mu\nu}\Box \phi,
\end{eqnarray}
and
\begin{eqnarray}\label{25}
\Box\phi=\alpha' T^{m\gamma}_{\gamma},
\end{eqnarray}
respectively,  where $\alpha'=\frac{8\pi}{2\omega_0+3}$ and $ T^{m\gamma}_{\gamma}=g^{\mu\nu}T^m_{\mu\nu}$ is the trace of the matter stress-tensor which becomes calculated from $L_m$ through the definition $T^m_{\mu\nu}=\frac{2}{\sqrt{-g}}\frac{\delta}{\delta g_{\mu\nu}}[\sqrt{-g}L_m]$.
For Bianchi type I spacetime, the field equations take the form
\begin{eqnarray}\label{26}
&&\frac{\phi^2}{4\omega_0}\left(3H^2-\sigma^2\right)-\frac{1}{2}\dot{\phi}^2
-\frac{3H}{2\omega_0}\phi\dot{\phi}=\rho_{\Lambda}+\rho_m,\\
&&\frac{-1}{4\omega_0}\left(2\dot{H}+3H^2+\sigma^2\right)\phi^2-\frac{1}{2}(1+\frac{1}{\omega_0})\dot{\phi}^2\cr
&&~~~~~~~~~~~~~~~~~~~~~~+\frac{H}{\omega_0}\phi\dot{\phi}-\frac{1}
{2\omega_0}\phi\ddot{\phi}=p_{\Lambda},\label{27}
\end{eqnarray}
and the wave equation is
\begin{eqnarray}\label{28}
\ddot{\phi}+3H\dot{\phi}-\frac{1}{2\omega_0}\left(3\dot{H}+6H^2+\sigma^2\right)\phi=0.
\end{eqnarray}
As above, Eqs.   (\ref{26}), (\ref{27}) and (\ref{28}),  are 3  independent equation which having unknown parameters such as $\phi$, $H$ and $\sigma$. To solve them  we take $\sigma^2=\sigma_0^2 H^2$ and $\phi=\phi _0a^{\epsilon}$  \citep{48}, where $\epsilon$ is any integer, implies that $\dot{\phi}=\epsilon H \phi$. So,  Eq.   (\ref{26})  lead to
\begin{eqnarray}\label{29}
1+2\epsilon-\frac{2}{3}\omega_0\epsilon^2-\frac{\sigma_0^2}{3}=\frac{4\omega_0}{3H^2\phi^2}(\rho_{\Lambda}+\rho_m).
\end{eqnarray}
The fractional energy densities are defined as
\begin{eqnarray}\label{30}
&&\Omega_m=\frac{\rho_m}{\rho_{cr}}=\frac{4\omega_0\rho_m}{3\phi^2H^2}, \cr
&&\Omega_{\Lambda}=\frac{\rho_\Lambda}{\rho_{cr}}=\frac{4\omega_0\rho_{\Lambda}}{3\phi^2H^2}=\frac{4\omega_0\alpha}{3\phi^2H},
\end{eqnarray}
where $\rho_{cr}=\frac{3\phi^2H^2}{4\omega_0}$. Therefore, Eq. (\ref{29}) give
\begin{eqnarray}\label{31}
\Omega_{\Lambda}+\Omega_{m}=1+2\epsilon-\frac{2}{3}\omega_0\epsilon^2-\Omega_{\sigma0}.
\end{eqnarray}
In the following, we take the time derivative of (\ref{29}), after using (\ref{31}), so
\begin{eqnarray}\label{32}
&&\frac{{H'(z)}}{H}=\frac{3}{2(1+z)}\cr
&&~~~~~~\times\left(1+\frac{2}{3}\epsilon +\frac{\Omega_{\Lambda}\omega_{\Lambda}}{1+2\epsilon-\frac{2}{3}\omega_0\epsilon^2-\Omega_{\sigma0}}\right).
\end{eqnarray}
For the case of  $\epsilon=0$, the above equation reduce to (\ref{16}).
Here by combining (\ref{8}) with (\ref{14}) and also (\ref{32}), we obtain the EoS parameter in BD theory as
\begin{eqnarray}\label{33}
&&\omega_{\Lambda}(z)=\frac{1+2\epsilon-\frac{2}{3}\omega_0\epsilon^2-\Omega_{\sigma0}}{-2(1+2\epsilon-\frac{2}{3}\omega_0\epsilon^2-\Omega_{\sigma0})+\Omega_{\Lambda}(z)}\cr
&&~~~~\times\bigg(1-\frac{2\epsilon}{3}+\frac{2  b^2(1+2\epsilon-\frac{2}{3}\omega_0\epsilon^2-\Omega_{\sigma0})}{\Omega_{\Lambda}(z)}\bigg).
\end{eqnarray}
The solar-system experiments give the lower bound for the value of $\omega_0$  to be $\omega_0>40000$ \citep{10}. However, when probing the larger scales, the limit obtained will be weaker than this result. It was shown \citep{49} that $\omega_0$ can be smaller than 40000 on the cosmological scales. Also, Sheykhi \textit{et al}. \citep{50} obtained  the result for the value of $\epsilon$ is $\epsilon<0.01$. The ghost DE model in BD framework has an interesting feature compared to the ghost DE model in BI Universe.
 In the case of     $b^2=0$, the EoS parameter of in the BD framework, requiring condition $\omega_{\Lambda}<-1$ leads to $(1+2\epsilon-\frac{2}{3}\omega_0\epsilon^2-\Omega_{\sigma0}) (3+2\epsilon) <3\Omega_{\Lambda}$. We can also obtain the evolution behavior of the DE. Taking the derivative of  (\ref{30})
as $\dot{\Omega}_{\Lambda}=-\Omega_{\Lambda}H(\frac{\dot{H}}{H^2}+2\epsilon)$ and  using relation $\dot{\Omega}_{\Lambda}=-H(z)(1+z)\Omega_{\Lambda}'(z)$,  it follows that
\begin{eqnarray}\label{34}
&&\Omega_{\Lambda}'(z)=-\frac{3\Omega_{\Lambda}}{1+z}\bigg(\frac{\Omega_{\Lambda}-1-2\epsilon+\frac{2}{3}\omega_0\epsilon^2+\Omega_{\sigma0}}{-2(1+2\epsilon-\frac{2}{3}\omega_0\epsilon^2-\Omega_{\sigma0})+\Omega_{\Lambda}(z)}\bigg)\cr
&&\bigg(1-\frac{2}{3}\epsilon+b^2(1+2\epsilon-\frac{2}{3}\omega_0\epsilon^2-\Omega_{\sigma0})\bigg).
\end{eqnarray}
Now,  the deceleration parameter   in BD theory is obtained as
\begin{eqnarray}\label{35}
&&q(z)=\frac{1}{2 }+\epsilon+\frac{3\Omega_{\Lambda}}{-2(1+2\epsilon-\frac{2}{3}\omega_0\epsilon^2-\Omega_{\sigma0})+\Omega_{\Lambda}(z)}\cr
&&~~~~~\times\bigg(1-\frac{2}{3}\epsilon+\frac{  b^2(1+2\epsilon-\frac{2}{3}\omega_0\epsilon^2-\Omega_{\sigma0})}{\Omega_{\Lambda}(z)}\bigg),
\end{eqnarray}
where $\Omega_{\Lambda}$ is given by Eq. (\ref{34}). A same steps as the pervious section can be followed to obtain the squared sound speed $c_s^2$  for Brans-Dicke theories. Taking time derivative of Eq. (\ref{33})  and replacing them into  the Eq. (\ref{20}) it is a matter of calculation to show that
\begin{eqnarray}\label{36}
&&c_s^2=\frac{\gamma}{3\Omega_\Lambda(2\gamma+\Omega_\Lambda)^2}\bigg[(2\gamma+\Omega_\Lambda)(6b^2 \gamma+\epsilon' \Omega_\Lambda)\cr
&&+\frac{(\gamma(\epsilon'+3b^2)+\epsilon'\Omega_\Lambda)(\epsilon'\Omega^2_\Lambda
+12\gamma b^2(\gamma+\Omega_\Lambda))}{\gamma(-3+3b^2-2\epsilon)-3\Omega_\Lambda}\bigg],
\end{eqnarray}
 \begin{figure}[tb]
 \includegraphics[width=.35\textwidth]{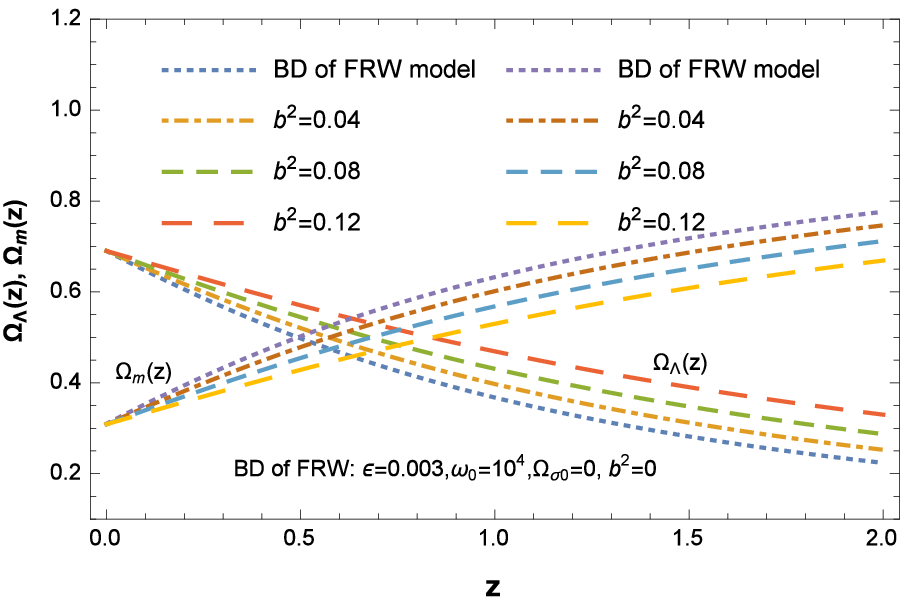}\hspace{.5cm}
 \includegraphics[width=.35\textwidth]{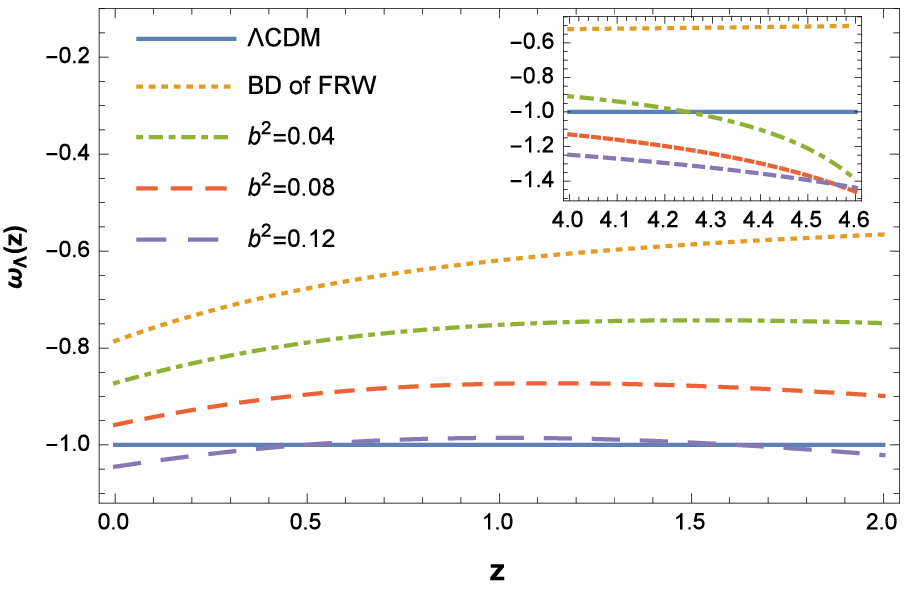}\hspace{.5cm}
 \includegraphics[width=.35\textwidth]{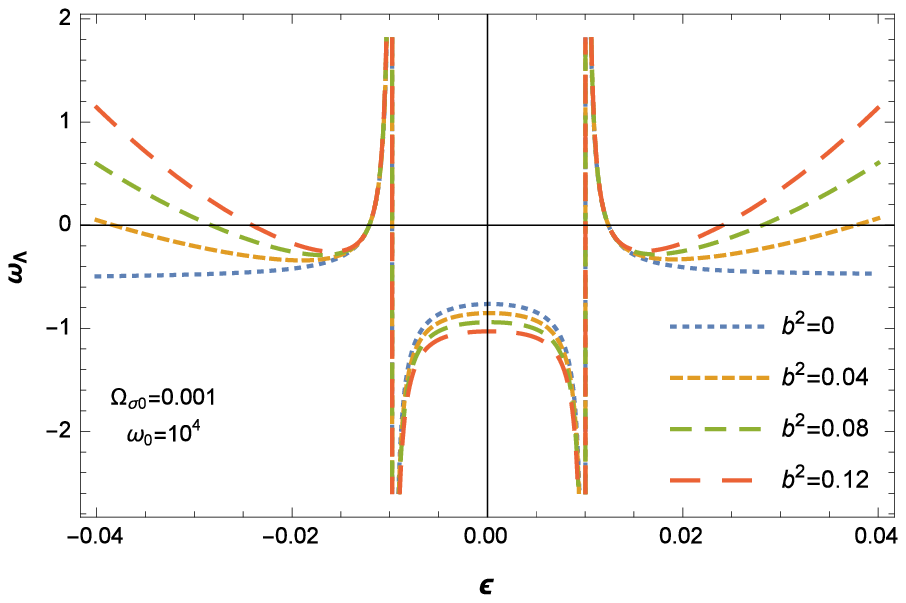}
\caption{Top panel: DE density parameters $\Omega_\Lambda$ and $\Omega_m$ for the interacting  ghost DE of BD theory with different $b^2$.  The evolutionary trajectories
of $\omega_\Lambda$ for the interacting ghost DE with $\epsilon=0.003$ with different values $b^2$  as a function of cosmic redshift $z$ (middle panel) and in terms  of $\epsilon$ (lower panel). Here we choose $\omega_0=10^4$, $\Omega^0_{\Lambda}=0.69$ and $\Omega_{\sigma0}=0.001$.}
 \label{fig:3}
 \end{figure}
 \begin{figure}[tb]
 \centerline{\includegraphics[width=.35\textwidth]{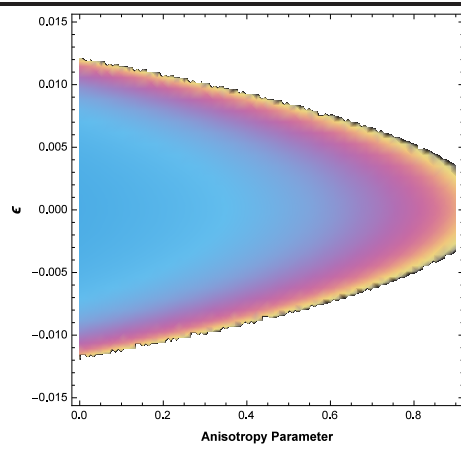}}
  \caption{The best fits of  $\epsilon$ with anisotropy parameter  for the interacting GDE model. The results given by current only are $z=1$,  $H_0=72~km/s/Mpc$,  $b^2=0.1$, and    $\Omega_{m0}=0.277$.}
 \label{fig:4}
 \end{figure}
where $\gamma=-1-2\epsilon+\frac{2}{3}\omega_0\epsilon^2 +\Omega_{\sigma0}$ and $\epsilon'=-3+2\epsilon$.  In Fig. 3 we plot the  energy density of DE component $\Omega_\Lambda(z)$ and the energy density of DM $\Omega_m(z)$ (upper panel), the redshift evolution of the equation of state $\omega_\Lambda(z)$ as a function of both  $z$  and $\epsilon$ in middle (lower panel) while the parameter $\epsilon$ versus the anisotropy parameter is plotted in figure (4).  In Fig. 5 we plot the   deceleration parameter   (upper panel) and  the squared sound speed (middle panel) as a function of the cosmic redshift $z$ for different parameter $b^2$ in BD theory. In the case of the   ghost DE of BD theory  we select the model parameter as  $\Omega_{\sigma0}=0.001$, $\Omega_{\Lambda}^0=0.69$, $\epsilon=0.003$ and  $\omega_{0}=10^4$. Fig. 3 (upper) indicates that at the late time $\Omega_\Lambda\rightarrow0.7$ while for the case of  the energy density of DM $\Omega_m\rightarrow0.3$, which is similar to the behaviour of the original  ghost DE in previous section.  From Fig. 3 (middle) we observe that for $b^2=\Omega_{\sigma0}= 0$,  the EoS parameter of BD theory translates
the Universe from low quintessence region  towards  high quintessence region. But   for  $b^2\neq0.12 $,  $\omega_{\Lambda}$ increases from phantom
region at early times  and approaches to  quintessence    region at late times. Also from Fig. 3 we see that  for  $b^2=0.12 $, $\omega_{\Lambda}$
of the interacting ghost DE in BD theory can cross the phantom divide and eventually the Universe approaches low phantom  phase of expansion  at late time. The
lower of   figure (3) indicates that one can generate a phantom-like behavior provided $-0.01<\epsilon<0.01$ which this point is completely compatible with the Ref. \citep{50}. For a better insight, we  plotted $\epsilon$ against the anisotropy parameter as shown in  figure 4. The sweet spot is estimated to be $z=1$.\\
We figure out that the behaviour of the deceleration parameter for the best-fit Universe is quite different from  the $\Lambda$CDM cosmology as shown in  Fig. 5  (upper panel). We can also see that the best fit values of transition
redshift and current deceleration parameter with ghost DE of BD theory are  $z=2.13^{+0.84+1.28}_{-0.00-0.55}$ and $q_0=-1.32^{+00+0.10}_{-0.07-0.17}$ which is matchable with the observations \citep{51} while for the case of $\Lambda$CDM, where $z\sim0.67$ and $q_0=-0.54$. We can
see that increasing $b^2$ decreases the value of $q(z)$.
 The evolution of $c_s^2$ against $z$ is plotted in Fig. 5 (middle panel) for different values of the coupling parameter $b^2$. The figure reveals that  $c_s^2$
is always negative and thus, as the previous case, a background filled with the interacting
ghost DE seems to be unstable against the perturbation. This implies that we cannot obtain a
stable  ghost DE dominated Universe in BD theory, which are in agreement with \citep{52,53}. One important point is the sensitivity of the instability to
the coupling parameter $b^2$. The larger $b^2$, leads to more instability against perturbations.\\
In the following, we study the capability of the $H(z)$ measurements in constraining DE models in BD theory.
The evolution of   Hubble parameter $H(z)$ in ghost DE model  with BD theory is obtained by using Eqs. (\ref{8}) and (\ref{29}) as follows
\begin{eqnarray}\label{37}
&&H=\frac{H_0}{-2\gamma} \bigg( -\Omega_{m0}-\gamma\cr
&&~~~~~~~~+\sqrt{(-\gamma-\Omega_{m0})^2- 4\gamma\Omega_{m0}(1+z)^{3}}\bigg).
\end{eqnarray}
   \begin{figure}[tb]
 \includegraphics[width=.35\textwidth]{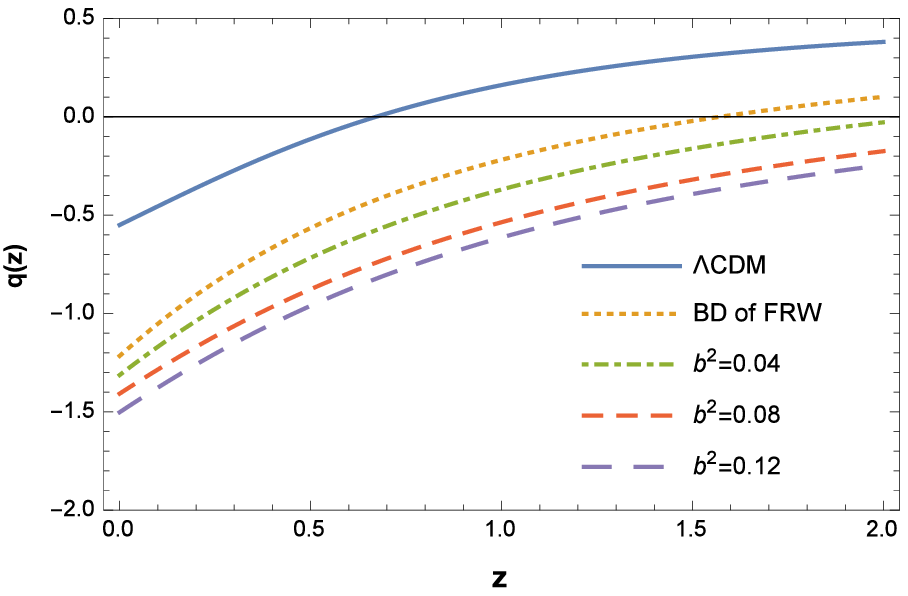}\hspace{.5cm}
 \includegraphics[width=.35\textwidth]{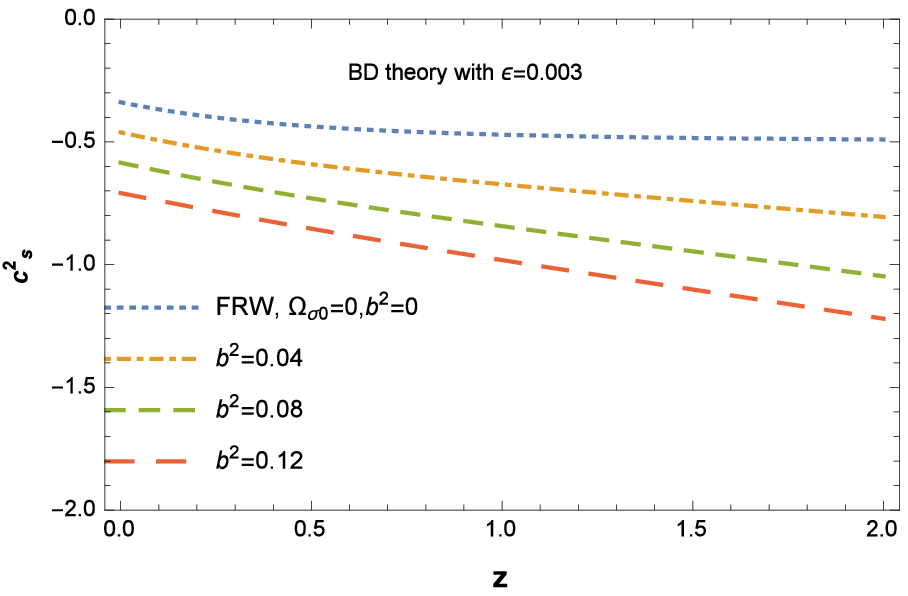}\hspace{.5cm}
 \includegraphics[width=.35\textwidth]{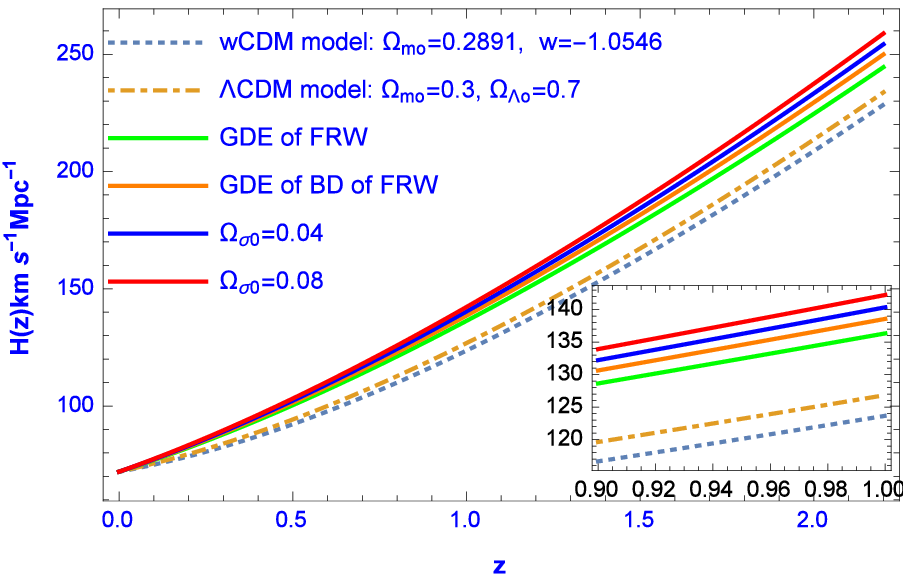}
\caption{Top panel: The evolution of $q(z)$ in terms of $z$  for the interacting ghost DE of BD theory with different $b^2$.   Middle panel: The evolution of  $c_s^2$ as a function of cosmic redshift $z$ for the different parameter $b^2$ with $\Omega^0_{\Lambda}=0.69$ and $\Omega_{\sigma0}=0.001$.   Lower panel: Hubble expansion parameter in terms of redshift for  the different parameter $\Omega_{\sigma0}$ with $b^2=0.1$, $\Omega_{m0}=0.277$,   $H_0=72kms^{-1}Mpc^{-1}$, $\epsilon=0.003$ and $\omega_0=10^4$. }
 \label{fig:5}
 \end{figure}
\begin{figure}[tb]
 \centerline{\includegraphics[width=.4\textwidth]{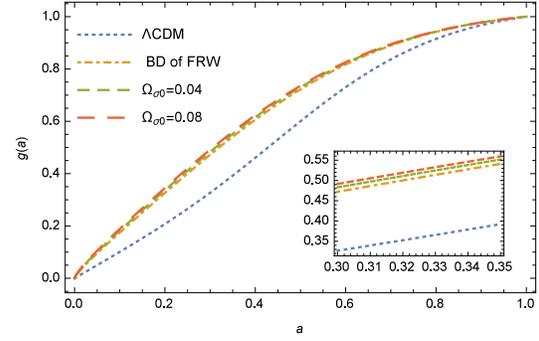}}
  \caption{ Evolution of growth function $g(a)$  in terms  of $a$ for the different $\Omega_{\sigma0}$   and comparing to the $\Lambda$CDM and FRW models in ghost DE of BD theory.}
 \label{fig:6}
 \end{figure}
 The behaviour of the Hubble parameter is   similar to that of the matter density parameters ($\Omega_m$), which is expected because DE comes to dominate the evolution
of the Hubble parameter only at very low redshift. We elect three specific DE models as representatives of cosmological models in order to make the analysis. They are the $\Lambda$CDM, $w$CDM, ghost DE of  BD theory in BI (FRW) models.   We consider to use  the SGL+CBS (the strong gravitational lensing,  the  cosmic microwave background, baryon acoustic oscillations and type Ia supernova)  data to constrain the $w$CDM and ghost DE  models and we take $\Omega_{m0}=0.2891$ and $w=-1.0546$ \citep{54}. As a matter of fact we can also see  the lower panel of Fig. 5 that in a BI model although ghost DE model performs a little poorer
than $\Lambda$CDM model, but it performs better than  ghost DE in  BD theory. Also, from this figure we can understand   the Hubble parameter  in  ghost DE of BD theory in BI are bigger than in  the  ghost DE of FRW,  $\Lambda$CDM and $w$CDM models.  The larger the Hubble expansion rate $H(z)$
 is taken, the bigger the anisotropy parameter $\Omega_{\sigma0}$  can reach. Therefore, from the
above analysis, we will figure out that both the parameters, $b^2$ and  $\Omega_{\sigma0}$, can impact the cosmic expansion history in the
interacting ghost DE of BD theory in BI model. \\
In Fig. (6) we show the effects of  anisotropy on  the growth factor   in ghost DE of BD theory for the DE models considered in this work, as compared to the $\Lambda$CDM model.  Generally,  the     $\Lambda$CDM model   observe less growth compared to the ghost DE of BD theory in an anisotropic Universe. Therefore the growth factor $g(a)$  for the  $\Lambda$CDM Universe   will always fall behind the  ghost DE models.\\
The theoretical distance modulus $\mu_{th}(z)$ is defined as \citep{55}
\begin{equation}\label{38}
\mu_{th}(z)=5 \log_{10}\frac{d_L(z)}{Mpc}+25,
\end{equation}
where $d_L(z)=(1+z)\int{ ^z_0 H^{-1}(z')}dz'$ is the luminosity distance. The structure of the anisotropies of the CMB radiation depends on two eras in cosmology, such as
last scattering and today. We can also  measure   $d_L(z)$  through the Hubble parameter by using the  Eq. (\ref{13}). Figure 7 presents the distance modulus with the best fit of our model and the best fit of the $\Lambda$CDM model. From Fig. (7) we can observe the Universe is accelerating expansion. In all, current data are unable to discriminate between the popular $\Lambda$CDM, FRW  and our interaction models.
\begin{figure}[tb]
 \centerline{\includegraphics[width=.4\textwidth]{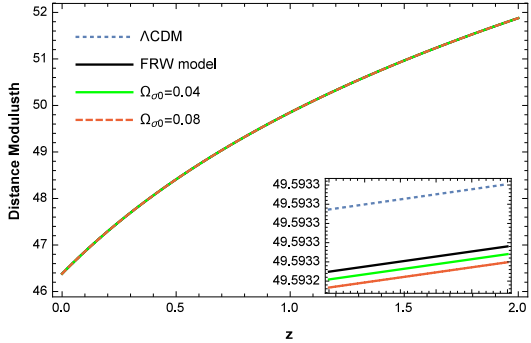}}
  \caption{ Distance modulus   for the best fit model $\Omega_{m0}=0.277$, $H_0=72~km/s/Mpc$,  $b^2=0.1$, and   the $\Lambda$CDM model,  $\Omega_{m0}=0.3$, $H_0=72~km/s/Mpc$ and $\Omega^0_{\Lambda}=0.7$.}
 \label{fig:7}
 \end{figure}

\section{Conclusion}
In this work we studied the linear evolution of structure formation in  interacting ghost DE models within the framework of Brans-Dicke theory.
first of all, we initiate our analysis by studying the effects of  anisotropy on the background expansion history of the growth factor.
We obtained the evolution of  density parameter $\Omega_\Lambda$, the equation of state parameter $\omega_\Lambda$,  the deceleration parameter $q$ and the  squared sound speed $c_s^2$ for both the ghost DE and Brans-Dicke theory  with respect to the cosmic redshift function. At first, the EoS parameter
of the ghost DE and BD theory models in the case of $b^2\neq0.12$, cannot cross the phantom divide while it for $b^2=0.12$  can cross the phantom divide line.  Beside, increasing of the anisotropy and the interaction parameter is increased the phantomic.
Then, the evolution of the interacting ghost DE  density parameter in BD theory is depend on the anisotropy density parameter $\Omega_{\sigma0}$ and the coupling constant $b^2$. On the basis of the above considerations, it seems reasonable to investigate an anisotropic Universe, in which the present
cosmic acceleration is followed by a decelerated expansion in an early matter dominant phase. In other words, it indicates that the values of transition scale factor and current
deceleration parameter are   $z=0.74^{+0.40+0.78}_{-0.00-0.28}$ and $q_0=-0.37^{+00+0.08}_{-0.09-0.19}$ for
the case of  ghost DE, $z=2.13^{+0.84+1.28}_{-0.00-0.55}$ and $q_0=-1.32^{+00+0.10}_{-0.07-0.17}$ for
the case of  ghost DE with BD theory while for the case of $\Lambda$CDM model, $z=0.67$ and $q_0=-0.54$  which is consistent  with   observations \citep{43,53}.
We have used the squared sound speed $c_s^2$ as the main factor to study the stability of the ghost DE in BD theory. As a result, a BI Universe filled with DM  and  ghost DE component in BD gravity can lead to an unstable  interacting ghost DE dominated Universe. In this case the frequency of the oscillations becomes purly imaginary and the density perturbations will grow with time.\\
Then, we analyzed $H(z)$ and compare the results with observational data. We found that, by choosing appropriate values of constant parameters, we figure out our model has more agreement with observational data than $\Lambda$CDM. Furthermore, we show that in anisotropic Universe with   ghost DE of BD theory,  the Hubble parameter are
bigger  than the  ghost DE of FRW,  $\Lambda$CDM and $w$CDM models.   It was
observed that the larger the Hubble expansion rate $H(z)$  is taken, the bigger the anisotropy parameter $\Omega_{\sigma0}$  can reach.
Finally   the effects of anisotropy  on the growth of structures in linear regime  is investigated and
we compared the linear growth in the ghost DE and BD theory  with the linear growth in the FRW   and   $\Lambda$CDM models which in the $\Lambda$CDM,
the growth factor evolves more slowly compared to the ghost DE of FRW in BD theory   because the cosmological constant dominates in the late time universe.  Also, in   the ghost DE of FRW in BD theory, the growth factor evolves more slowly
compared to the    ghost DE models in anisotropic Universe.  Therefore  due to BD theory the growth factor $g(a)$ for the $\Lambda$CDM
Universe will always fall behind the ghost DE models in an anisotropic Universe.


\end{document}